\documentclass[aps,pra,twocolumn,superscriptaddress,showpacs,showkeys,a4paper,english]{revtex4-1}
\usepackage{amsmath,epsfig,amssymb,subfigure,bm,dsfont}
\usepackage{times} \usepackage{graphicx} \usepackage{float}
\usepackage{color} \usepackage{subfigure} \usepackage{epstopdf}
\usepackage[colorlinks=true,linkcolor=blue,citecolor=blue]{hyperref}
\usepackage{hyperref} \usepackage{soul} \usepackage{ulem}
\usepackage{bbold} \usepackage[english]{babel}
\usepackage[OT1]{fontenc} \synctex=1 \usepackage[T1]{fontenc}
\usepackage{graphicx} \DeclareGraphicsExtensions{.png,.jpg}
\graphicspath{{fig/}}

\newcommand\be{\begin{equation}} \newcommand\ee{\end{equation}}
\newcommand\bea{\begin{eqnarray}} \newcommand\eea{\end{eqnarray}}
\newcommand\bmul{\begin{multline}} \newcommand\emul{\end{multline}}
\newcommand\bes{\begin{subequations}}
  \newcommand\esu{\end{subequations}}

\sethlcolor{yellow}

\begin{document}

\title{An efficient perturbation theory of density matrix
  renormalization group}

\author{Emanuele Tirrito} \author{Shi-Ju~Ran} \author{Andrew J.
  Ferris} \affiliation{ICFO - Institut de Ciencies Fotoniques, The
  Barcelona Institute of Science and Technology, 08860 Castelldefels
  (Barcelona), Spain} \author{Ian P. McCulloch} \affiliation{Centre
  for Engineered Quantum Systems, School of Physical Sciences, The
  University of Queensland, Brisbane, Queensland 4072, Australia}
\author{Maciej Lewenstein} \affiliation{ICFO - Institut de
  Ciencies Fotoniques, The Barcelona Institute of Science and
  Technology, 08860 Castelldefels (Barcelona), Spain}
\affiliation{ICREA - Instituci\'{o} Catalana de Recerca i Estudis
  Avan\c{c}ats, Lluis Companys 23, 08010 Barcelona, Spain}

\begin{abstract}
  Density matrix renormalization group (DMRG) is one of the most
  powerful numerical methods available for many-body systems.
  It has been applied to solve many physical problems, including
  calculating ground-states and dynamical properties. In this work, we
  develop a perturbation theory of DMRG (PT-DMRG) to largely increase
  its accuracy in an extremely simple and efficient way. By using the
  canonical matrix product state (MPS) representation for the ground
  state of the considered system, a set of orthogonal basis functions
  $\left\lbrace \vert \psi_i \rangle \right\rbrace$ is introduced to
  describe the perturbations to the ground state obtained by the
  conventional DMRG. The Schmidt numbers of the MPS that are beyond
  the bond dimension cut-off are used to define such perturbation
  terms. The perturbed Hamiltonian is then defined as $\tilde{H}_{ij}=
  \langle \psi_i \vert \hat{H} \vert \psi_j \rangle$; its ground state
  permits to calculate physical observables with a considerably
  improved accuracy as compared to the original DMRG results. We
  benchmark the second-order perturbation theory with the help of
  one-dimensional Ising chain in a transverse field and the Heisenberg chain, where the
  precision of DMRG is shown to be improved $\rm O(10)$ times.
  Furthermore, for moderate $L$ the errors of DMRG and PT-DMRG both 
  scale linearly with $L^{-1}$ (with $L$ being the length of the chain). 
  The linear relation between the dimension cut-off of DMRG and that of PT-DMRG
  with the same precision shows a considerable improvement of
  efficiency, especially for large dimension cut-off's. In thermodynamic limit we
  show that the errors of PT-DMRG scale with $\sqrt{L^{-1}}$. Our work suggests an 
  effective way to define the tangent space of the ground state MPS, which may shed lights on the properties beyond the ground
  state. Such second-order PT-DMRG can be readily generalized to higher orders, 
  as well as applied to the models in higher dimensions.
\end{abstract}

\maketitle
\section{Introduction}
In the last three decades, strongly-correlated quantum many-body
systems remain in the center of scientific interests and define the
most important challenges and open questions
\cite{S11,LMM11,CMS12,LSA12}. For instance, understanding of certain
class of quantum many-body systems is necessary for the understanding
of the mechanism of high-$T_c$ superconductivity (cf.
\cite{A87,LNW06}), or of topological phase transitions (cf.
\cite{KT72, BH13}) and spin liquids (cf. \cite{LB10}, for the recent
experiment see \cite{BBY16}). These systems are notoriously hard to be
studied analytically or numerically. Exact solutions are extremely
rare for such kind of systems. In fact, the Bethe ansatz works well
only for one dimensional systems (cf.
\cite{bethe_ansatz_1,bethe_ansatz_2, EFGK05}). In various mean field
theories, the role of quantum fluctuations is usually underestimated.
For these reasons, novel efficient numerical approaches are highly
desired. These new approaches naturally encounter great challenges,
because the dimension of Hilbert space of considered systems increases
exponentially with number of particles. This limits significantly not
only the applicability of exact diagonalization methods \cite{AWS10},
but even quantum Monte Carlo methods \cite{TW05}; the latter can be
applied for larger systems, but they face the fatal negative sign
problem for fermionic and frustrated systems.

One of the most important numeric tool developed in the last decades
is the method based on tensor networks \cite{US11,RO14}.  It offers an
efficient representation of quantum many-body states that coincides
with their entanglement structure. It takes advantage of the fact that
not all quantum states in the Hilbert space of many-body systems with
(in particular short-range interactions) are equally relevant for the
low-energy and low-temperature physics. It has been found namely that
the low-lying eigenstates of gapped Hamiltonians with local
interactions obey the so-called area law of the entanglement entropy
\cite{area_law_ECP,VLRK03,LRV,PKL99,JDB73,MS93,PEDC05,CC05}.
Specifically speaking, for a spatial subregion $\mathcal{R}$ of the
physical space where the system is defined, the reduced density matrix
is defined as $\hat{\rho}_{\mathcal{R}}= \rm Tr_{\mathcal{E}}
(\hat{\rho})$, with $\mathcal{E}$ denoting the spatial complement of
the subregion $\mathcal{R}$. The entanglement entropy is defined as
\be
  S(\rho_{\mathcal{R}}) = - \rm Tr \lbrace \rho_{\mathcal{R}} \rm
  log (\rho_{\mathcal{R}} ) \rbrace .  
\ee
Then the area law of the entanglement entropy reads 
\be
  S(\rho_{\mathcal{R}})= \rm O(\vert \partial \mathcal{R} \vert) , 
\ee
with $\vert \partial \mathcal{R}
\vert$ the length of the boundary. In particular for a $D$-dimensional
lattice, one has 
\be
S=O(L^{D-1})
\ee
with $L$ being the length scale.
This means that for one-dimensional (1D) systems, $S= \rm
\textit{const}$. The area law suggests that the low-lying eigenstates
stay in a ``small corner'' of the full Hilbert space of the many-body
system, and that they can be described by a much smaller number of
parameters. This subset of states can be well approximated by tensor
network states.

The density matrix renormalization group (DMRG)
\cite{DMRGRev1,DMRGRev2} is one of the most famous tensor network
methods, based on the so-called matrix product state (MPS), a
one-dimensional (1D) TN state ansatz \cite{US11}. DMRG algorithm was
formulated by S. White in 1992 for calculating ground state properties
of 1D strongly-correlated systems \cite{dmrg_white1,dmrg_white2}. The
original DMRG is a variant of Wilson's numeric renormalization group
\cite{NRG} with Hilbert space decimations and reduced basis
transformations. Instead of truncating the eigenstates of Hamiltonian
according to their energies, the selection is based on their weights
in the reduced density matrices, i.e. the entanglement. Such a
strategy improves the performance largely. It was then realized by S.
\"{O}stlund and S. Rammer that the block states in DMRG can be
represented as MPS \cite{OR95}, where they predicted the properties of
the entanglement spectrum, such as area law \cite{JDB73,MS93}. F.
Verstraete \textit{et al.} reinterpreted the DMRG algorithm as a
variational principle from the perspective of quantum information
theory \cite{dmrg_VPC}.

DMRG has extremely wide applications in 1D strongly-correlated
systems, e.g. for simulating ground state properties of 1D spin
\cite{B82} or Hubbard \cite{H63,H64,K63,G63,LSA12,LSABD07} chains.
Referring to the spin models, DMRG accurately gives the excitation gap
of the $S=1$ Heisenberg chain \cite{WH93}, or for Haldane gap
\cite{SA94,SA94_2}. DMRG shows also a great efficiency when applied
for fermionic systems, such as 1D Hubbard model and t-J model
\cite{JS07}, where logarithmic corrections to the correlations were
found, as compared with $S=1/2$ Heisenberg chain. Moreover, DMRG has
been used to study the topological order and quantum Hall effect
\cite{SJ95, TM99}.

DMRG has also been extended to two-dimensional (2D) models
\cite{SW12}, and one of the most remarkable achievements of DMRG is
the demonstration of the quantum spin liquid behavior in 2D frustrated
magnets that break no symmetries even down to zero temperature
\cite{LB10}. By calculating topological entanglement entropy
\cite{KP06}, strong evidence for a spin liquid ground state was found
using DMRG for the Heisenberg antiferromagnet on kagome lattice
\cite{YHW11}. DMRG has also been used to identify spin liquid phases
stabilized by anisotropic next to next neighbour, and multi-spin
interactions \cite{WSWB06,JWS11,CSW,ShouShu,GZBS}. But, 2D DMRG
suffers from finite-size effects, and thus the definitive evidence for
the existence of isotropic spin models with short range interactions
\cite{JKMSZW09,SMF09,BSMF,WNS94}, whose ground states break no
symmetries, is still missing.

The DMRG method was also developed to the study of dynamic properties,
such as dynamical structure functions or frequency-dependent
conductivities \cite{H95,SPKSB,KW99,J02}. At the same time, its
finite-temperature extensions to 2D classical \cite{N95} and 1D
quantum \cite{WX97,S97} systems show good performance and precision.
It has even been utilized to more demanding study of non-Hermitian
(pseudo-) Hamiltonians emerging in the analysis of the relaxation
towards classical steady states in 1D systems far from equilibrium
\cite{H98,CHS99,HS01}.

In this paper we develop a perturbation theory of DMRG (PT-DMRG) that
provides a remarkably efficient way to improve the precision of DMRG.
We define a set of states forming an orthogonal basis $\lbrace \vert
\psi_i \rangle \rbrace$, obtained from the conventional DMRG. The
perturbed Hamiltonian is then defined as $\tilde{H}_{ij} = \langle
\psi_i \vert \hat{H} \vert \psi_j \rangle$. The ground state of
$\tilde{H}$ permits to calculate physical observables with a
considerably improved accuracy as compared to the original DMRG
results. We test our method on the quantum Ising model in a transverse
field and on Heisenberg model. In particular, we show how the error committed by DMRG and
PT-DMRG scales with the bond dimension $\chi$ and the length of chain
$L$. Without increasing the computation costs much, the error is
reduced about $O(10)$ times using PT-DMRG. Other perturbation scheme
are explained in \cite{lepetit98, hubing15, dukelsky98, jutho13}.

This paper is organized as follows: In section II, we briefly review
DMRG and present some discussion about its convergence properties. In
sections III and IV, we describe the PT-DMRG and discuss its
properties. In section V, we discuss the numerical results on the
quantum transverse Ising model. In section V a summary and an outlook
are presented.

\section{Density Matrix Renormalization Group}

\begin{figure}
  \includegraphics[angle=0,width=0.75\linewidth]{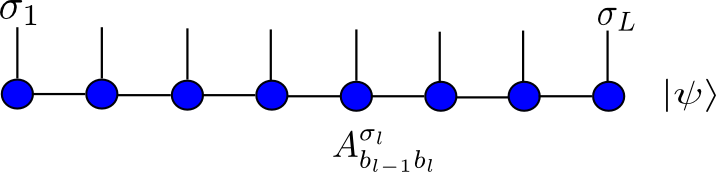}
  \caption{Graphical representation of matrix product state (MPS).}
  \label{mps0}
\end{figure}

Let us consider a 1D quantum system consisting of $L$ sites. Each
lattice site has physical degrees of freedom denoted as $| \sigma_j
\rangle$ in a local $d$-dimension Hilbert space
$\mathcal{H}_d=\mathcal{C}^d$. A pure state can be generally written
in a local basis as
\be
  \vert \psi \rangle = \sum_{\sigma_1 \ldots
    \sigma_L} C_{\sigma_1 \ldots \sigma_L} | \sigma_1 \ldots \sigma_L
  \rangle.  
\ee
with $C_{\sigma_1 \ldots \sigma_N}$ the coefficient matrix. If the
lattice has open boundary condition, $C_{\sigma_1 \ldots \sigma_N}$
can be rewritten in an MPS using a series of singular value
decomposition (SVD) as 
\be
\label{eq::mps} \vert \psi
\rangle = \sum_{\sigma_1 \ldots \sigma_L} A^{\sigma_1}_{1,b_1}
A^{\sigma_2}_{b_1,b_2} \ldots A^{\sigma_L}_{b_{L-1},1} \vert
\sigma_1 \ldots \sigma_L \rangle, 
\ee
where
$A^{\sigma_i}_{\beta_{i-1},\beta_i}$ is a third-order tensor, i.e., a
($\chi_{i-1} \times \chi_i$) matrix for each value of $\sigma_i$, with
$\chi_i$ the bond dimension of the index $b_i$ [Fig. \ref{mps0}]. The
state represented in Eq. (\ref{eq::mps}) is called as matrix product
state (MPS).

Considering the Hamiltonian $\hat{H}$ with nearest-neighbor
interactions 
\be
\hat{H} = \sum_{l=1}^{L-1}
\hat{H}_{l,l+1}.  
\ee
In order to obtain the ground state $\vert \psi_0 \rangle$, one needs
to find the MPS that minimizes the following equation 
\be
  E= \frac{\langle \psi \vert \hat{H} \vert \psi \rangle}{\langle \psi
    \vert \psi \rangle}.  
\ee
The most efficient way of
doing this is in a variational approach by minimizing $E$ over MPS
family 
\be
\min_{\vert \psi \rangle \in MPS} \left\lbrace \langle \psi
  \vert \hat{H} \vert \psi \rangle -\lambda \langle \psi \vert \psi
  \rangle \right\rbrace .
\ee
Ideally, the minimization should be done simultaneously over all the
coefficients of all tensors. However, this is quite difficult and
inefficient to implement. Following the original procedure\cite{dmrg_white1,dmrg_white2}, 
the strategy of DMRG that we use here is to minimize two
tensors each time while keeping others fixed. Then, we move to another
pair of tensors and repeat the procedure until convergence. 
In detail, defined $D^{\sigma_l \sigma_{l+1}}_{\alpha
  \beta} = \sum_{\gamma} A^{\sigma_l}_{\alpha, \gamma}
A^{\sigma_{l+1}}_{\gamma, \beta}$ as the contraction of the two
unfixed tensors. Then the minimization is written as
\begin{multline}
  \label{eq:min1}
  \min_{\vert \psi \rangle \in MPS} \left\lbrace \langle \psi \vert \hat{H} \vert \psi \rangle -\lambda \langle \psi \vert \psi \rangle \right\rbrace  \rightarrow \\
  \min_{D} \left\lbrace D^{\dagger} \hat{H}_{eff} D -\lambda
    D^{\dagger} \hat{N} D \right\rbrace.
\end{multline}
$\hat{H}_{eff}$ and $\hat{N}$ correspond to $\langle \psi \vert
\hat{H} \vert \psi \rangle$ and $\langle \psi \vert \psi \rangle$
without $D$ and $D^{\dagger}$, respectively. The term $-\lambda
\langle \psi \vert \psi \rangle$ is introduced to make all eigenvalues
negative, so that MPS is generated to converge to the ground state. By
considering $D$ as a vector, the minimization becomes
\be\label{eq:min2} \frac{\partial}{\partial D^{\dagger}}
  \left\lbrace
    D^{\dagger} \hat{H}_{eff} D -\lambda D^{\dagger} \hat{N} D
  \right\rbrace =0 .
\ee
To proceed, we introduce two vectors 
\bea
\label{eq:alpha_beta_1}
\vert \alpha \rangle &=& \sum_{\sigma_1 \ldots \sigma_{l-1}} \left( A^{\sigma_1} \ldots A^{\sigma_{l-1}} \right)_{1,\alpha} \vert \sigma_1 \ldots \sigma_{l-1} \rangle , \\
\vert \beta \rangle &=& \sum_{\sigma_{l+2} \ldots \sigma_N} \left(
  A^{\sigma_{l+2}} \ldots A^{\sigma_N} \right)_{\beta,1} \vert
\sigma_{l+2} \ldots \sigma_N \rangle .
\label{eq:alpha_beta_2}
\eea
Then the state $\vert \psi \rangle$ can be written as follows
\be
\label{eq::ffdec} \vert \psi \rangle = \sum_{\sigma_l
  \sigma_{l+1}
  \alpha \beta} D^{\sigma_l \sigma_{l+1}}_{\alpha \beta} \vert \alpha
\sigma_l \sigma_{l+1} \beta \rangle .  
\ee
\begin{figure}
  \includegraphics[scale=0.70]{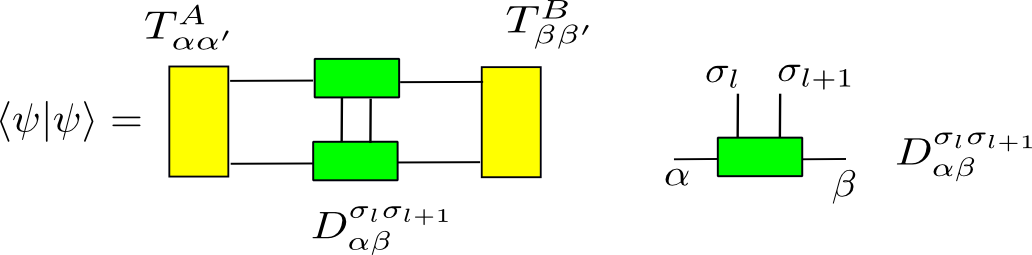}
  \caption{Graphical representation of $\langle \psi \vert \psi
    \rangle$ through the 2-rank tensors $T^{A}$ and $T^{B}$.}
  \label{mps1}
\end{figure}
Let us first consider the overlap $\langle \psi \vert \psi \rangle$.
As shown in Fig. \ref{mps1}, we use the Eq. (\ref{eq::ffdec}) 
\be
\langle \psi \vert \psi \rangle = \sum_{\sigma_l \sigma_{l+1}}
\sum_{\alpha \alpha'} \sum_{\beta \beta'} T^{A}_{\alpha \alpha'}
D^{\sigma_l \sigma_{l+1}}_{\alpha \beta} {D^{\sigma_l
    \sigma_{l+1}}_{\alpha' \beta'}}^{\dagger} T^{B}_{\beta \beta'},
\ee
where $T^{A}_{\alpha \alpha'}$ and $T^{B}_{\beta\beta'}$ are 
\bea
T^{A}_{\alpha \alpha'} &=& \sum_{\sigma_1 \ldots \sigma_{l-1}} \left( {A^{\sigma_{l-1}}}^{\dagger} \ldots {A^{\sigma_1}}^{\dagger} A^{\sigma_1} \ldots A^{\sigma_{l-1}}  \right)_{\alpha \alpha'}, \\
T^{B}_{\beta \beta'} &=& \sum_{\sigma_{l+2} \ldots \sigma_N} \left(
  A^{\sigma_{l+2}} \ldots A^{\sigma_N} {A^{\sigma_N}}^{\dagger} \ldots
  {A^{\sigma_{l+2}}}^{\dagger} \right)_{\beta \beta'}.  
\eea

\begin{figure}
  \includegraphics[scale=0.70]{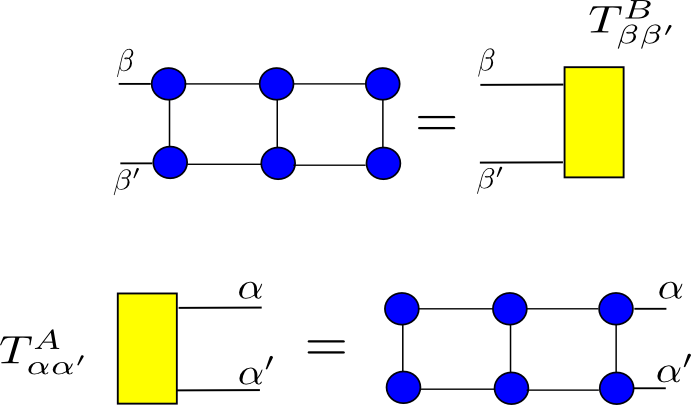}
  \caption{Graphical representation of the matrices $T^{A}$ and
    $T^{B}$ that contain the contraction.}
  \label{ta_tb}
\end{figure}

The tensor $\hat{T}^A$ ($\hat{T}^B$) contains all the contraction of
tensors of MPS from site $1$ to site $l-1$ ($l+2$ to $L$) (see Fig.
\ref{ta_tb}).  If the basis from the site $1$ to $l-1$ are
left-orthogonal and the basis from $l+1$ to $N$ are right-orthogonal,
we simply have
\begin{align}
  T^{A}_{\alpha \alpha'} = \delta_{\alpha \alpha'}, \quad T^{B}_{\beta
    \beta'} = \delta_{\beta \beta'} .
\end{align}
We will show below that such left- and right- orthogonal conditions
are automatically fulfilled in DMRG.

\begin{figure}
  \includegraphics[scale=0.6,width=0.85\linewidth]{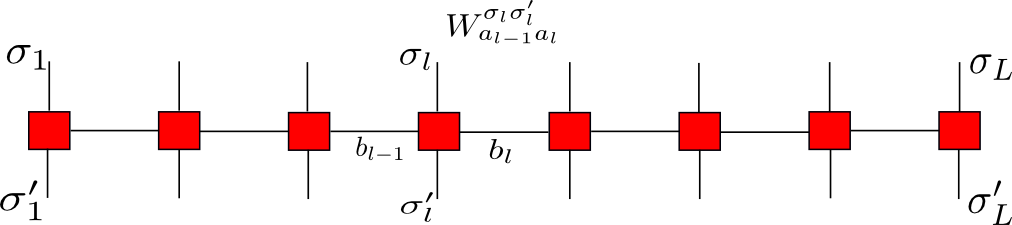}
  \caption{Matrix product operator representation of $\hat{H}$. In
    each site is defined an 4-rank tensor $\hat{W}^{\sigma_l
      \sigma'_l}_{a_{l-1} a_l}$.}
  \label{mpo}
\end{figure}

Let us now consider the quantity $\langle \psi \vert \hat{H} \vert
\psi \rangle$. Assume that we can write $\hat{H}$ in a matrix product
operator (MPO) \cite{CDV08,ND13,IPM} (Fig. \ref{mpo}), i.e., 
\be
\hat{H} = \hat{W}^{[1]}_{1,b_1} \hat{W}^{[2]}_{b_1, b_2} \ldots
\hat{W}^{[L]}_{b_{L-1}, 1}, 
\ee
where $\hat{W}^{[l]} = \sum_{\sigma_l \sigma'_l} W^{\sigma_l \sigma'_l} \vert \sigma_l
\rangle \langle \sigma'_l \vert $ is defined in a local Hilbert space.

\begin{figure}
  \includegraphics[width=1.0\linewidth]{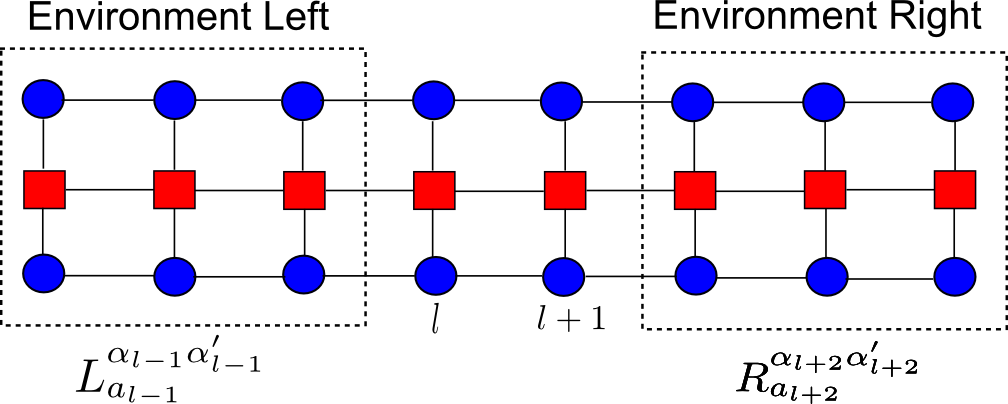}
  \caption{Tensor network represented the quantity $\langle \psi \vert
    \hat{H} \vert \psi \rangle$.}
  \label{psiH}
\end{figure}

The $\langle \psi \vert \hat{H} \vert \psi \rangle$ is described in
the tensor network in the Fig. \ref{psiH} that containing the
contraction between two MPS and the MPO. Therefore, one has
\begin{multline}
  \langle \psi \vert \hat{H} \vert \psi \rangle = \sum_{\sigma_l \sigma_{l+1} \alpha \beta} \sum_{\sigma'_l \sigma'_{l+1} \alpha' \beta'} {D^{\sigma_l \sigma_{l+1}}_{\alpha \beta}}^{\dagger} D^{\sigma'_l \sigma'_{l+1}}_{\alpha' \beta'} \cdot \\
  \langle \alpha \sigma_l \sigma_{l+1} \beta \vert \hat{H} \vert
  \alpha' \sigma'_l \sigma'_{l+1} \beta' \rangle .
\end{multline}
Let us now look at the matrix elements $\langle \alpha \sigma_l
\sigma_{l+1} \beta \vert \hat{H} \vert \alpha' \sigma'_l \sigma'_{l+1}
\beta' \rangle$ using the MPO representation of Hamiltonian $H$
\begin{multline}
  \label{eq:meH}
  \langle \alpha \sigma_l \sigma_{l+1} \beta \vert \hat{H} \vert \alpha' \sigma'_l \sigma'_{l+1} \beta' \rangle = \sum_{\sigma \sigma'} W^{\sigma_1 \sigma'_1} \ldots W^{\sigma_L \sigma'_L} \\
  \cdot \langle \alpha \sigma_l \sigma_{l+1} \beta \vert \sigma_1
  \ldots \sigma_L \rangle \langle \sigma'_1 \ldots \sigma'_L \vert
  \alpha' \sigma'_l \sigma'_{l+1} \beta' \rangle .
\end{multline}
Using the equations (\ref{eq:alpha_beta_1}) and
(\ref{eq:alpha_beta_2}), we can evaluate the scalar product in the
previous equation
\begin{multline}
  \langle \alpha \sigma_l \sigma_{l+1} \beta \vert \sigma'_1 \cdots \sigma'_L \rangle = \left( {A^{\sigma_1}}^{\dagger} \cdots {A^{\sigma_{l-1}}}^{\dagger} \right)_{1,\alpha} \\
  \cdot \left( {A^{\sigma_{l+2}}}^{\dagger} \cdots
    {A^{\sigma_L}}^{\dagger} \right)_{\beta, 1},
\end{multline}

\begin{multline} 
  \langle \sigma'_1 \cdots \sigma'_L  \vert \alpha' \sigma'_l \sigma'_{l+1} \beta' \rangle = \left( A^{\sigma'_1} \cdots A^{\sigma'_{l-1}} \right)_{1,\alpha'}  \\
  \cdot \left( A^{\sigma'_{l+2}} \cdots A^{\sigma'_L} \right)_{\beta',
    1} .
\end{multline}

\begin{figure}
  \includegraphics[scale=0.60]{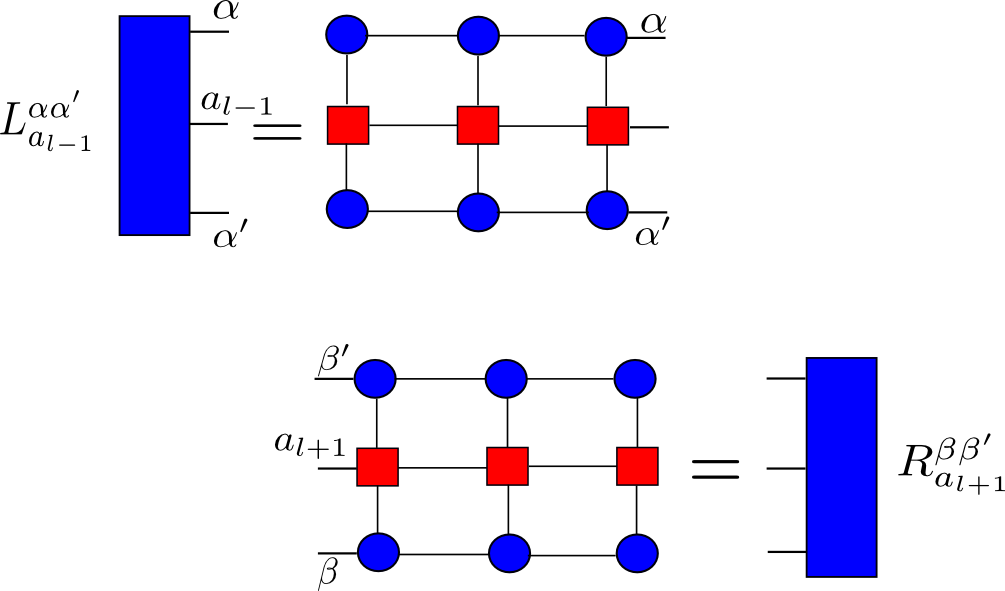}
  \caption{Graphical representation of environment left $L$ and right
    $R$, where $L$ contain the contracted left part while $R$ contain
    the contracted right part of network.  }
  \label{env}
\end{figure}

Define the tensors $L$ and $R$ that contain the contracted left and
right halves as (see Fig. \ref{env})
\begin{multline}
  \label{eq:EL}
  {L}^{\alpha \alpha'}_{a_{l-1}} = \left\lbrace \sum_{\sigma_1 \sigma'_1} {A^{\sigma_1}_{1,b_1}}^{\dagger} W^{\sigma_1 \sigma'_1}_{1, a_1} A^{\sigma'_1}_{1,b'_1} \right\rbrace \cdots \\
  \cdots \left\lbrace \sum_{\sigma_{l-1} \sigma'_{l-1}}
    {A^{\sigma_{l-1}}_{b_{l-2},b_{l-1}}}^{\dagger} W^{\sigma_{l-1}
      \sigma'_{l-1}}_{a_{l-2}, a_{l-1}}
    A^{\sigma'_{l-1}}_{b'_{l-2},b'_{l-1}} \right\rbrace ,
\end{multline}
\begin{multline}
  \label{eq:ER}
  {R}^{\beta \beta'}_{a_{l+1}} = \left\lbrace \sum_{\sigma_{l+2} \sigma'_{l+2}} {A^{\sigma_{l+2}}}^{\dagger} W^{\sigma_{l+2} \sigma'_{l+2}}_{a_{l+1} a_{l+2}} A^{\sigma'_{l+2}} \right\rbrace \cdots \\
  \cdots \left\lbrace \sum_{\sigma_L \sigma'_L}
    {A^{\sigma_L}_{b_{l+1}, b_{l+2}}}^{\dagger} W^{\sigma_L
      \sigma'_L}_{a_{L-1} 1} A^{\sigma'_L}_{b'_{l+1}, b'_{l+2}}
  \right\rbrace.
\end{multline}
Through the Eqs. (\ref{eq:EL}) and (\ref{eq:ER}), we obtain
\begin{multline}
  \langle \alpha \sigma_l \sigma_{l+1} \beta \vert \hat{H} \vert \alpha' \sigma'_l \sigma'_{l+1} \beta' \rangle = \\
  =\sum_{a_{l-1} a_l a_{l+1}} {L}^{\alpha \alpha'}_{a_{l-1}}
  W^{\sigma_l \sigma'_l}_{a_{l-1} a_l} W^{\sigma_{l+1}
    \sigma'_{l+1}}_{a_l a_{l+1}} {R}^{\beta \beta'}_{a_l} .
\end{multline}

\begin{figure}
  \includegraphics[scale=0.70]{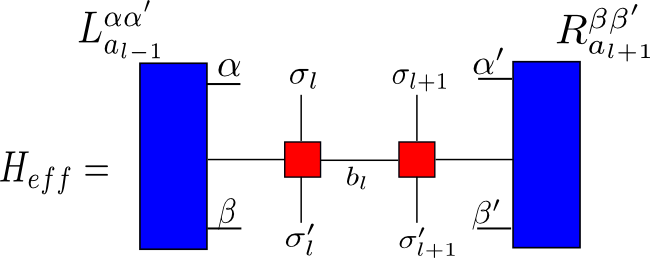}
  \caption{Graphical representation of effective Hamiltonian $H_{eff}$
    defined in equation (\ref{heff}). }
  \label{mps}
\end{figure}

Now we can immediately write $\langle \psi \vert \hat{H} \vert \psi
\rangle$ as
\begin{multline}
  \langle \psi \vert \hat{H} \vert \psi \rangle = \sum_{\alpha \alpha'} \sum_{\beta \beta'} \sum_{\sigma_l \sigma'_l} \sum_{\sigma_{l+1} \sigma'_{l+1}} {D^{\sigma_l \sigma_{l+1}}_{\alpha \beta}}^{\dagger} D^{\sigma_l \sigma_{l+1}}_{\alpha' \beta'} \cdot \\
  \cdot {L}^{\alpha \alpha'}_{a_{l-1}} W^{\sigma_l \sigma'_l}_{a_{l-1}
    a_l} W^{\sigma_{l+1} \sigma'_{l+1}}_{a_l a_{l+1}} {R}^{\beta
    \beta'}_{a_l} ,
\end{multline}
and rewrite Eq. (\ref{eq:min2}) as
\begin{multline}
  \sum_{\alpha' \beta'} \sum_{\sigma_l \sigma'_l} \sum_{\sigma_{l+1} \sigma'_{l+1}} L^{\alpha \alpha'}_{a_{l-1}} W^{\sigma_l \sigma'_l}_{a_{l-1} a_l} W^{\sigma_{l+1} \sigma'_{l+1}}_{a_l a_{l+1}} R^{\beta \beta'}_{a_{l+1}} D^{\sigma_l \sigma'_l}_{\alpha' \beta'}  \\
  - \lambda \sum_{\alpha' \beta} T^{A}_{\alpha \alpha'} T^{B}_{\beta
    \beta'} D^{\sigma_l \sigma_{l+1}}_{\alpha \beta} =0 .
\end{multline}
The matrices $H_{eff}$ (see Fig. \ref{mps}) and $N$ simply are 
\be
\label{heff} H_{eff} = \sum_{a_{l-1} a_l a_{l+1}} L^{\alpha
  \alpha'}_{a_{l-1}} W^{\sigma_l \sigma'_l}_{a_{l-1} a_l}
W^{\sigma_{l+1} \sigma'_{l+1}}_{a_l a_{l+1}} R^{\beta
  \beta'}_{a_{l+1}} , 
\ee
\be 
N =T^{A}_{\alpha \alpha'} T^{B}_{\beta \beta'}.
\ee

Using the expressions above, the minimization problem becomes 
\be
H_{eff} D - \lambda N D =0.  
\ee
After solving $D^{\sigma_l\sigma_{l+1}}_{\alpha \beta}$, 
we update $A^{\sigma_l}_{\alpha \beta}$
by performing a SVD
\be
D^{\sigma_l \sigma_{l+1}}_{\alpha
  \beta} = \sum_{\rho} U_{\sigma_l \alpha, \rho} S_{\rho} V_{\rho,
  \sigma_{l+1} \beta}.
\ee
Take only the $\chi$ largest 
singular vectors in $U$ as the new tensor $A^{\sigma_l}_{\alpha
  \rho}$, i.e., $A^{\sigma_l}_{\alpha \rho} = U_{\sigma_l \alpha,
  \rho}$ when sweeping from left to right, and take the $\chi$ largest
singular vectors in $V$ as the new tensor
$A^{\sigma_{l+1}}_{\rho,\beta}$ when sweeping from right to left. In
this way, the left and right orthogonal conditions of the MPS are
automatically fulfilled.

\begin{figure}
  \includegraphics[scale=0.70]{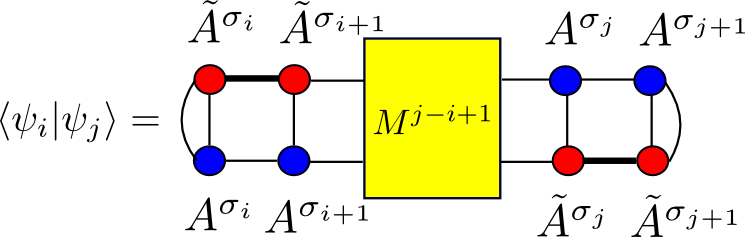}
  \caption{Graphical representation of the overlap $\langle \psi_i |
    \psi_j \rangle$ represented in the equation (\ref{eq:eq_ovl_pt}).}
  \label{ovl_pert}
\end{figure}


Specifically speaking, a left-to-right (or right-to-left) sweep
consists of the following steps:
\begin{itemize}
\item Start with a random initial MPS and transform it in the right
  orthogonal form.
\item Optimize the tensor $D^{\sigma_i \sigma_{i+1}}$: construct the
  environment $L$ and $R$ and solve the standard eigenvalue problem:
  \be H_{eff} D - \lambda D=0
    \label{eq:eigensolver}
  \ee
\item Carry out an SVD of $D^{\sigma_i \sigma_{i+1}}$ and update the
  tensor $A^{\sigma_i}$.
\item Repeat the same operations for every site until reaching the
  preset convergence: 
\be
\langle \psi \vert \hat{H}^2 \vert \psi
\rangle - \left(\langle \psi \vert \hat{H} \vert \psi \rangle
\right)^2 \rightarrow 0.
\ee
\end{itemize}
To analyse the computational cost we have to take special care to
ensure optimal ordering of multiplications when dealing with each
eigensolver given by (\ref{eq:eigensolver}). The problem is to
contract $L_{i-1} W_i W_{i+1} R_{i+2} D^{\sigma_i \sigma_{i+1}}$, with
$L_{i-1}$ $R_{i+2} \in (\chi, \chi_W, \chi)$, $W_i \in (d,\chi_W, d
\chi_W) $ and $D^{\sigma_i \sigma_{i+1}} \in (\chi, d, \chi)$. The
optimal ordering should be $(((L_{i-1} D^{\sigma_i \sigma_{i+1}})W_i
W_{i+1})R_{i+2}$, and in the way, one has to
\begin{itemize}
\item Contract $L_{i-1}$ and $D^{\sigma_i \sigma_{i+1}}$ over the left
  MPS bond at a cost $O(\chi^3 \cdot \chi_W \cdot d^2)$.
\item Multiply with $W_i W_{i+1}$ over the physical bond of $D^{i
    i+1}$ at a cost $O(\chi^2 \cdot \chi_W^2 \cdot d^4)$.
\item Contract with $R_{i+2}$ over the right MPO and MPS bond at a
  cost $O(\chi^3 \cdot d^2 \cdot \chi_W)$
\end{itemize}
The total cost of this procedure to apply $\hat{H}$ to $\vert \psi
\rangle$ is $O(\chi^3 \cdot \chi_W \cdot d^2 + \chi^2 \cdot \chi_W^2
\cdot d^4 + \chi^3 \cdot d^2 \cdot \chi_W)$.
\section{Subspace Expansion}
In the following, we develop a second-order perturbation theory for
DMRG. Note that from the orthogonality that the contribution of the
first-order term is zero. This optimization permits the recovery of
some of the lost information due to the truncation in the SVD of 
$D^{\sigma_i \sigma_{i+1}}$, and reach a better approximation of the ground
state.  In last section, we have shown how DMRG works and where its
error comes from. To reduce the error, we define a new orthogonal
basis $\lbrace \vert \psi_i \rangle \rbrace$, whose elements have the
MPS form. We put an impurity bond in each $\lbrace \vert \psi_i
\rangle \rbrace$ so that it is orthogonal to the ground state obtained
by DMRG. To define this impurity bond (e.g. between the $i$-th and
($i+1$)-th sites), we consider the SVD of $D^{\sigma_i \sigma_{i+1}}$
and the tensor $\tilde{A}^{\sigma_i}$ as the second $\chi$ largest
singular vectors. Thus, $\tilde{A}^{\sigma_i}$ is orthogonal to the
tensor $A^{\sigma_i}$ in the original MPS.

By introducing one impurity in different bonds of $\vert \psi_0
\rangle$, we can define a new basis $\lbrace \vert \psi_i \rangle
\rbrace$. Since both are in orthogonal form, one has
\begin{multline}
  \langle \psi_i \vert \psi_j \rangle  = \sum A^{\sigma_{i}} \left( \tilde{A}^{\sigma_{i}} \right)^{\dagger} A^{\sigma_{i+1}} \left( \tilde{A}^{\sigma_{i+1}} \right)^{\dagger}  M^{j-i+1}  \\
  \tilde{A}^{\sigma_j} \left( A^{\sigma_j} \right)^{\dagger}
  \tilde{A}^{\sigma_{j+1}} \left( A^{\sigma_{j+1}} \right)^{\dagger},
  \label{eq:eq_ovl_pt}
\end{multline}
where $M$ is the transfer matrix of the overlap $\langle \psi_i |
\psi_j \rangle$ (see Fig. \ref{ovl_pert}). Thus, $|\psi_i \rangle$ and
$|\psi_j \rangle$ are orthogonal to each other for $i \neq j$.

Now one can define the perturbed Hamiltonian $\hat{\mathcal{H}}$ with
$\lbrace \vert \psi_i \rangle \rbrace$ ($i=0, 1, \cdots$). Note that
$| \psi_0 \rangle$ is the ground state by the original DMRG. The
matrix elements of $\hat{\mathcal{H}}$ are defined as
\be
\mathcal{H}_{ij} = \langle \psi_i \vert \hat{\mathcal{H}} \vert \psi_j
\rangle
\label{eq-Hpt}
\ee
and form the matrix $\mathcal{H}$. The ground state
energy is calculated as 
\be
\tilde{E}_0 = \frac{\langle
  \tilde{\psi}_0 \vert \hat{\mathcal{H}} \vert \tilde{\psi}_0
  \rangle}{\langle \tilde{\psi}_0 \vert \tilde{\psi}_0 \rangle}
\ee
where $\vert \tilde{\psi}_0 \rangle$ is defined as the
combination of $\lbrace \vert \psi_i \rangle \rbrace$
\be
  \vert \tilde{\psi}_0 \rangle = \sum_j \Psi_j \vert \psi_j \rangle,
\ee
where $\Psi_j$ are the coordinates of the dominant
eigenvector of $\hat{\mathcal{H}}$. By using that the basis $\lbrace
\vert \psi_i \rangle \rbrace$, the perturbed ground state energy is
simply obtained as 
\be
\tilde{E}_0 = \sum_{i j}
\Psi^{\dagger}_{j}\mathcal{H}_{ij} \Psi_j.  
\ee

\section{Perturbation Theory DMRG}

Now we explain how to implement the PT-DMRG in practice. Using the
notation introduced above, the steps follow mostly the standard DMRG.
In an outermost loop, the update sweeps over the system from left to
right and right to left until the preset convergence is reached. The
inner loop sweeps over the system, iterating over and updating the
tensors on each site sequentially. Each local update during a left to
right sweep consists of the following steps:
\begin{itemize}
\item Perform the standard DMRG to obtain the ground state MPS
  $|\psi_0 \rangle$ (which is assumed in the right-orthogonal form).
\item From left to right, calculate $D^{\sigma_i \sigma_{i+1}}$ and
  perform SVD for each $i$; Keep the second $\chi$ largest left and
  right singular vectors as $\tilde{A}^{\sigma_i}$ and
  $\tilde{A}^{\sigma_{i+1}}$, respectively.
\item Construct the orthogonal basis $\lbrace \vert \psi_i \rangle
  \rbrace$ for by putting an impurity $\tilde{A}^{\sigma_i}$ in
  different bonds.
\item Construct the perturbed Hamiltonian $\hat{\mathcal{H}}$
  according to Eq. (\ref{eq-Hpt}) and calculate its dominant
  eigenvector $\Psi$.
\item Calculate the perturbed ground state of the systems as
  \be 
  \vert \tilde{\psi}_0 \rangle = \sum_{i=1}^N \Psi_i
  \vert \psi_i \rangle .
  \ee
\end{itemize}
As regards the computational cost, in addition, we need to consider the diagonalization of $\hat{\mathcal{H}}$ 
in the subspace.
This cost is $O(N^3)$ where $N$ is the number of the perturbed basis.
Therefore the full cost is $\chi^3 d \chi_W + O(\chi^2)+ O(N^3)$,
which makes it quite expensive. But the diagonal and first row/column of 
$\cal{H}$ can be obtained easily during the final DMRG sweep itself, which
makes it much more practical.
\begin{figure}
  \includegraphics[scale=0.5,angle=0,width=0.855\linewidth]{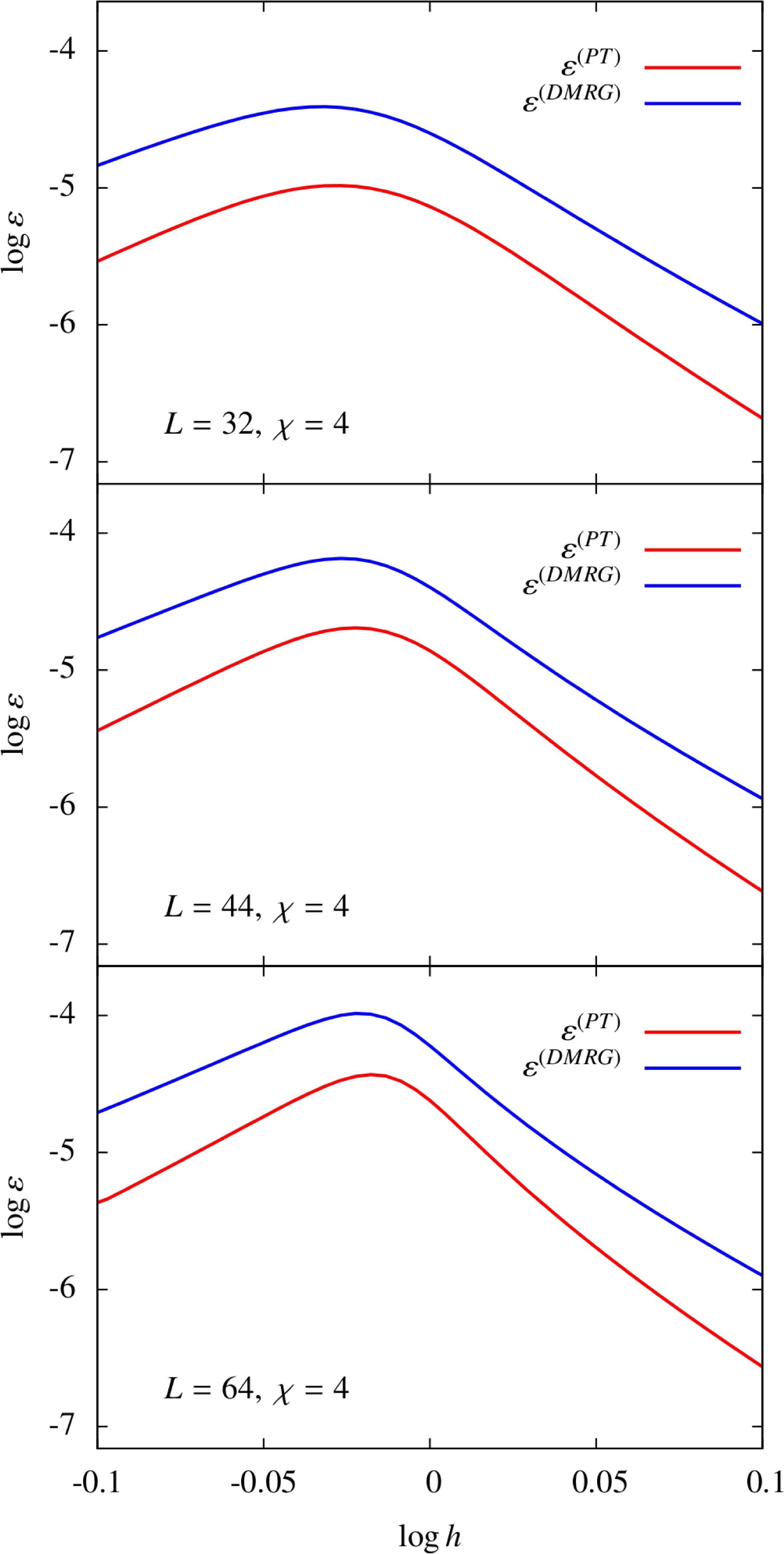}
  \caption{The error $\varepsilon$ of 1D Ising model on $32, 44, 64$
    chain with open boundary condition as a function of $h$. The error
    of PT-DMRG method with bond dimension $\chi=4$ is more than
    $O(10)$ time smaller compared with the error of standard DMRG.}
  \label{error_N}
\end{figure}

\section{Numerical Results}

\begin{figure}
  \includegraphics[scale=0.5,angle=0,width=0.88\linewidth]{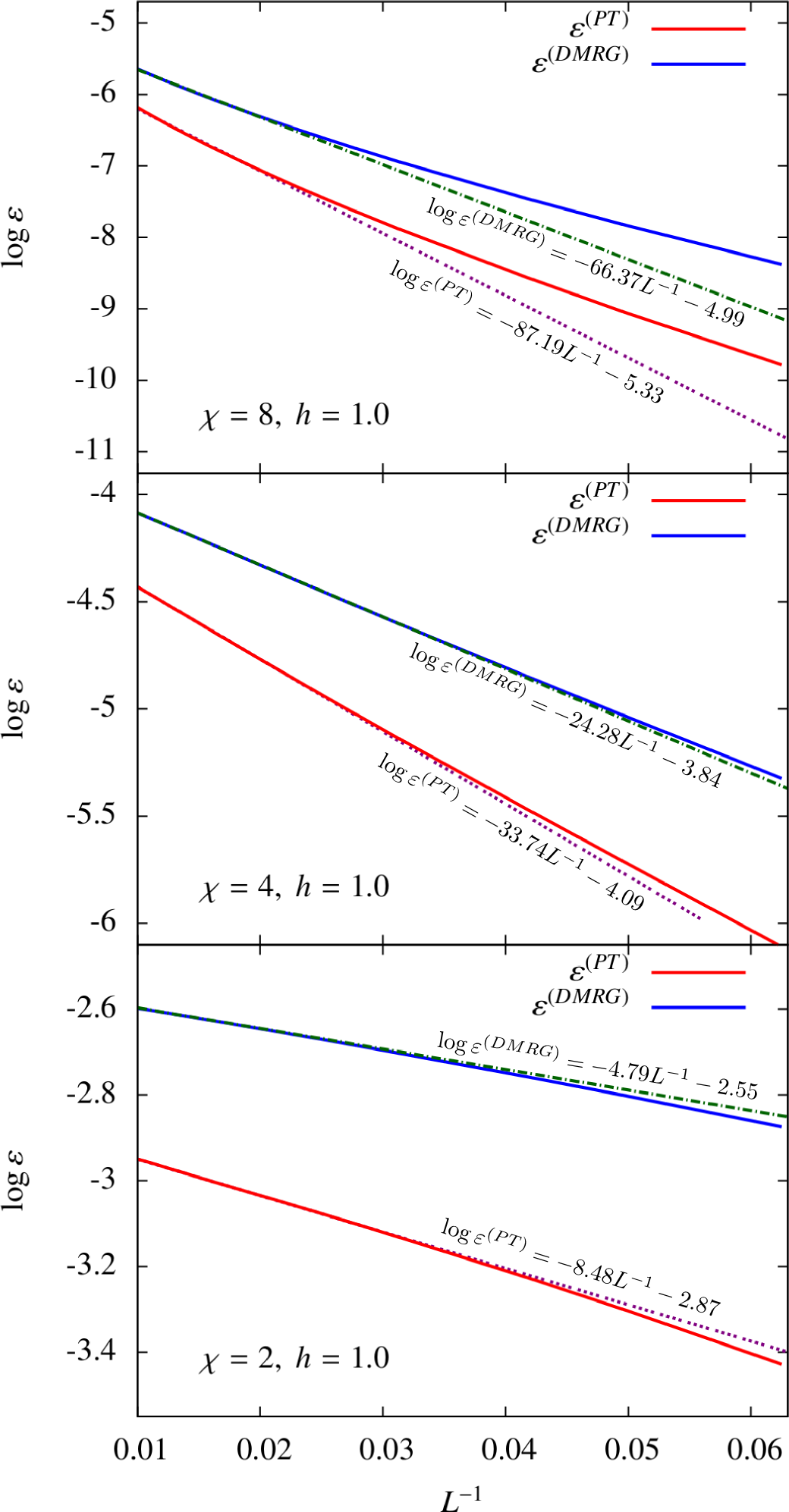}
  \caption{The error $\varepsilon$ of 1D Ising model as a function of
    length in the quantum phase transition $h=1.0$, for different
    values of $\chi=2,4,8$. We show that the error PT-DMRG is much
    smaller. The PT-DMRG gives a systematic improvement of accuracy. }
  \label{error_L}
\end{figure}

\subsection*{Quantum Transverse Ising Model}
To illustrate our method we study the 1D spin-half quantum Ising model
in a transverse field, especially near the quantum phase transition.
The Hamiltonian reads 
\be
\hat{H}= -J \sum_{i=1}^{L}
\hat{\sigma}^{x}_i \hat{\sigma}^{x}_{i+1} +h \sum_{i=1}^L
\hat{\sigma}^{z}_i .
\ee
In the infinite case, a quantum
phase transition occurs at $h/J=1$. The system for $h/J>1.0$ is in a
paramagnetic phase with an order parameter $\langle S^{x}\rangle\neq
0$, and in a ferromagnetic phase for $h/J<1.0$ with an order parameter
$\langle S^{z}\rangle\neq 0$. At the critical point, both order
parameters go to zero. We set $J=1$ as the energy scale.

In the numerical simulations, we considered a finite-size system with
open boundary condition with the length $L=16 \sim 128$. To benchmark
PT-DMRG, we compute the ground state energies of DMRG and PT-DMRG with
the same bond dimension $\chi$, and compare with the (quasi-exact)
result from the DMRG with sufficiently large $\chi=100 \sim 400$ (note
$\chi$ for quasi-exact calculations changes according with the length
of the chain, in other words the entanglement). The error is defined as 
\be
\varepsilon = \frac{E_0-\langle \psi \vert \hat{H} \vert \psi \rangle}{E_0},
\ee
with $E_0$ the energy from the quasi-exact DMRG.



\begin{figure}
  \includegraphics[scale=0.5,angle=0,width=0.98\linewidth]{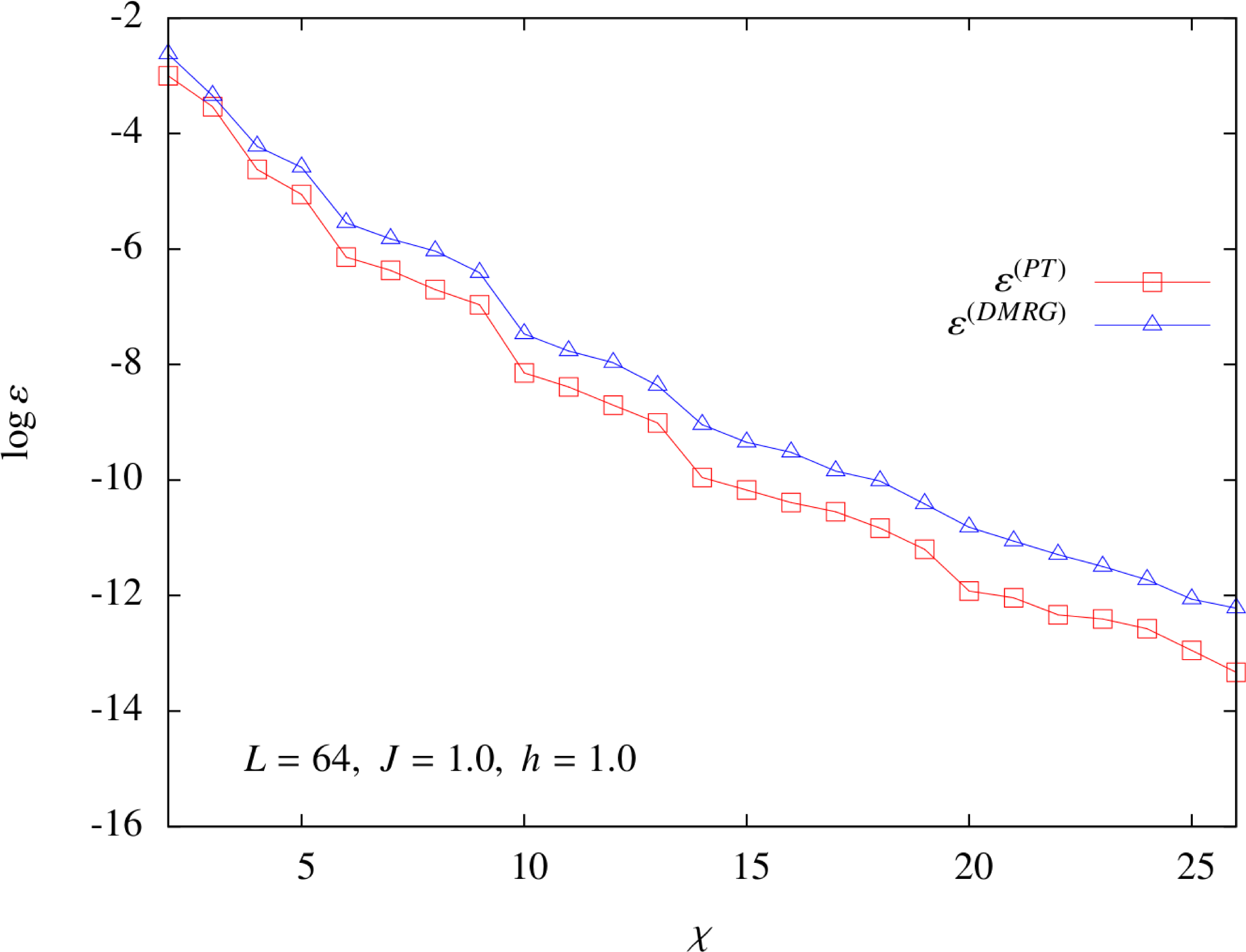}
  \caption{The error $\varepsilon$ of 1D Ising model versus of $\chi$
    in the quantum phase transition $h=1.0$ for $L=64$. We show how
    the error of PT-DMRG decrease faster than the error of the
    standard DMRG.}
  \label{error_chi_ising}
\end{figure}

The Fig. \ref{error_N} shows the error with $L=32,44,64$ versus
magnetic field $h$. We compare the results of the conventional DMRG
and PT-DMRG for $\chi=2,4,8$. Near to the phase transition, the error
of PT-DMRG is more than $O(10)$ times smaller compared with the error
of the conventional DMRG with the same $\chi$. Our simulations suggest
that through PT-DMRG, we are able to retrieve the leading term of the
lost information with the truncations in the SVD.

In Fig. \ref{error_L}, we show the error against $L^{-1}$ for $h=1$
(critical point). The results show that the error increase both
linearly with $L^{-1}$ for DMRG and PT-DMRG, indicating a systematic
improvement of the accuracy for moderate values of $L$. For the thermodynamic
limit the error of PT-DMRG scales as $\sqrt{L^{-1}}$, for reasons 
explained below. 

In Fig. \ref{error_chi_ising}, we show the error against $\chi$ for
$h=1$ (phase transition) and for $L=64$. The results show that the
error decrease with bond dimension $\chi$ for DMRG and PT-DMRG. The
error of PT-DMRG decreases faster than that of standard DMRG. This
shows considerable improvement of the accuracy for any value of bond
dimension $\chi$ near the phase transition.
\begin{figure}
  \includegraphics[scale=0.5,angle=0,width=1.0\linewidth]{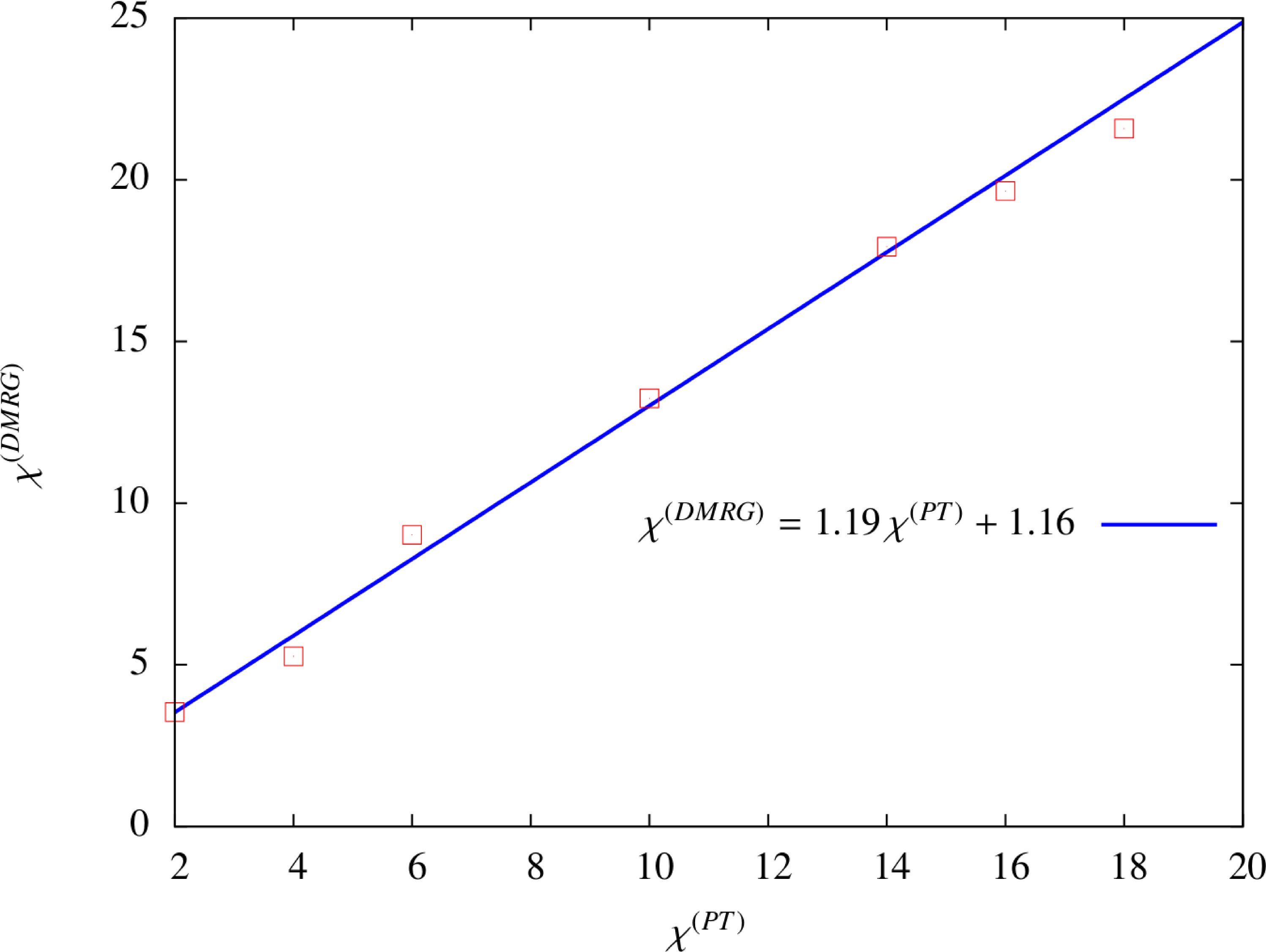}
  \caption{The plot is the fit between $\chi^{(DMRG)}$ versus of
    $\chi^{(PT)}$ of 1D Ising model in the quantum phase transition
    $h=1.0$ for $L=64$. We show how the PT-DMRG needs smaller bond
    dimension $\chi$ than DMRG.}
  \label{pt_vs_dmrg_ising}
\end{figure}

To see more clearly the improvement of the efficiency of PT-DMRG, we
study the correspondence between the bond dimension cut-off
$\chi^{(DMRG)}$ of the standard DMRG and that of PT-DMRG
$\chi^{(PT)}$. As shown in Fig. \ref{pt_vs_dmrg_ising}, each pair of
$\chi^{(DMRG)}$ and $\chi^{(PT)}$ given by the data points
approximately have the same precision. In detail, to determine
$\chi^{(DMRG)}$ for a given $\chi^{(PT)}$, we first find two $\chi$'s
with DMRG, where the precision of one $\chi$ is higher than the
precision of PT-DMRG with $\chi^{(PT)}$, and the other is lower. Then,
we do a fit to find $\chi^{(DMRG)}$, which is an fraction between
these two $\chi$'s.

We choose $h=1$ and $L=64$. The results show that with each
$\chi^{(PT)}$ in PT-DMRG, we need a larger bond dimension cut-off
(i.e. keep more states) in DMRG to reach the same precision. We also
find a linear relation between $\chi^{(PT)}$ and $\chi^{(DMRG)}$ as
\be
\chi^{(DMRG)} = 1.19 \chi^{(PT)} + 1.16.
\ee
Since the computational cost an MPS takes scales as
$\sim \chi^2$ (2 is the number of the virtual bond in each local
tensor of MPS), such a linear relation suggests that the larger $\chi$
one uses, the more computational resource one can save by using
PT-DMRG.


\subsection*{Heisenberg Model}

We study also the 1D spin-half quantum Heisenberg model, where the
Hamiltonian reads 
\be
\hat{H}= -J \sum_{i=1}^{L} (\hat{\sigma}^{x}_i \hat{\sigma}^{x}_{i+1}
+ \hat{\sigma}^{y}_i \hat{\sigma}^{y}_{i+1} +\hat{\sigma}^{z}_i
\hat{\sigma}^{z}_{i+1}).
\ee
We take $J=1$ as energy scale.

\begin{figure}
  \includegraphics[scale=0.5,angle=0,width=0.98\linewidth]{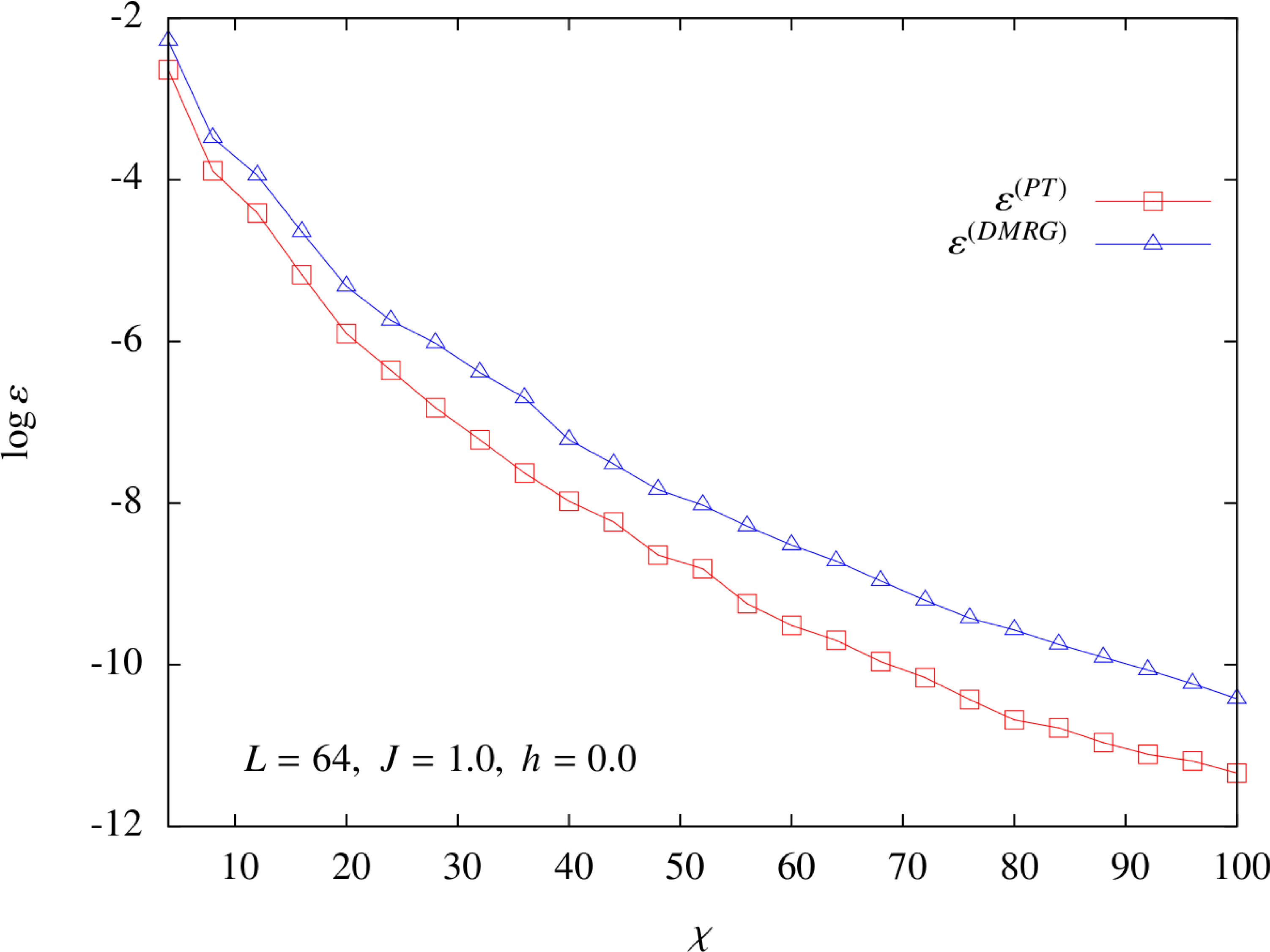}
  \caption{The error $\varepsilon$ of 1D Heisenberg model versus of
    $\chi$ in the quantum phase transition $h=0.0$ for $L=64$. We show
    how the error of PT-DMRG decrease faster than the error of the
    standard DMRG.}
  \label{error_chi_heisenberg}
\end{figure}
In Fig. \ref{error_chi_heisenberg}, we show the error against $\chi$
for $L=64$. The results show that the error decrease with bond
dimension $\chi$ for DMRG and PT-DMRG. Amazingly, the error of PT-DMRG
decreases faster than that of standard DMRG. This shows considerable
improvement of the accuracy for any value of bond dimension $\chi$.

\begin{figure}
  \includegraphics[scale=0.3,angle=0,width=0.97\linewidth]{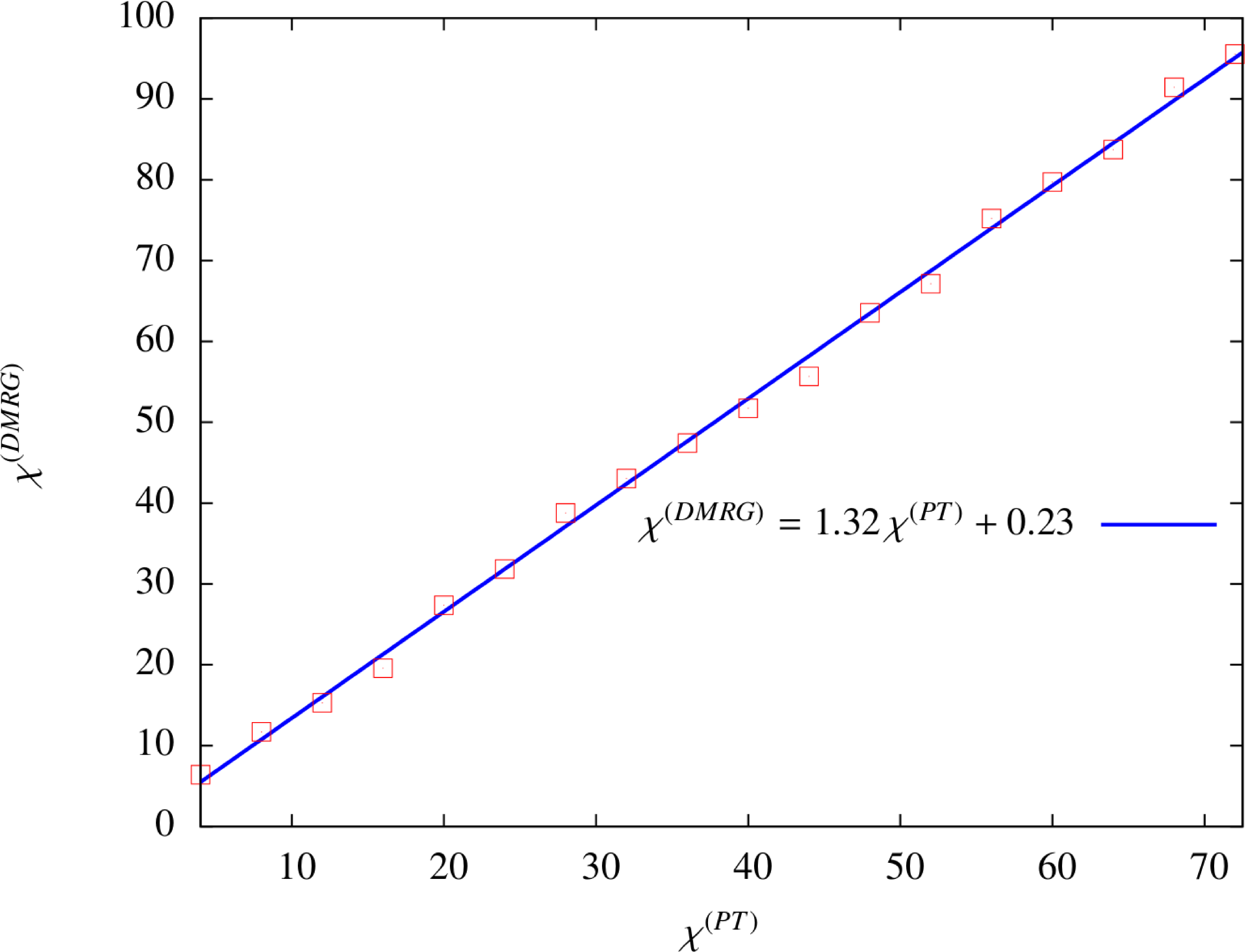}
  \caption{The plot is the fit between $\chi^{(DMRG)}$ versus of
    $\chi^{(PT)}$ of 1D Heisenberg model in the quantum phase
    transition $h=0.0$ for $L=64$. We show how the PT-DMRG needs
    smaller bond dimension $\chi$ than DMRG.}
  \label{pt_vs_dmrg_heisenberg}
\end{figure}

In Fig. \ref{pt_vs_dmrg_heisenberg}, we show the fit of
$\chi^{(DMRG)}$ against $\chi^{(PT)}$ for $L=64$. Again, a linear
relation is found between $\chi^{(PT)}$ and $\chi^{(DMRG)}$ as 
\be
\chi^{(DMRG)} = 1.32 \chi^{(PT)} + 0.23.
\ee
Especially,
the slope is larger than that in the quantum Ising model, which
implies a more significant improvement of efficiency when calculating
Heisenberg chain with a large bond dimension cut-off.
\section{Thermodynamic Limit}
In the following, we explore a second order perturbation theory for
DRMG in the thermodynamic limit. In the previous section we showed
that the error scaling of PT-DMRG is linear in $L^{-1}$ for moderate $L$. Now if $L$
approach to infinity we have that the scaling law is $1/\sqrt{L}$.

We focus on the results first in the Fig. \ref{error_L}. If we extend
the results to larger $L$ we can see a changing in behaviour for large
$L$ limit, the error in the energy per site becomes exactly equal to
that of conventional DMRG.

We can understand that from looking at how the PT-DMRG approaches the
thermodynamic limit. The off-diagonal matrix elements of effective
Hamiltonian $\mathcal{H}_{ij}$ for $ \vert i-j \vert > 1$ decay
exponentially quickly, so it really only needs a few of them. For the
Ising model $\mathcal{H}_{i,i+2}$ is already $O(10^{-6})$, so this
gives no improvement over the old style of calculating just the
diagonal part and the overlap with the ground state.  In the large $L$
limit, the effective Hamiltonian $ \mathcal{H}_{ij}$ can be
well-approximated by: 
\be 
\mathcal{H} =\begin{bmatrix}
  a & b & b & b & b &\cdots \\
  b & c & 0 & 0 & 0 &\cdots \\
  b & 0 & c & 0 & 0 &\cdots  \\
  b & 0 & 0 & c & 0 & \cdots \\
  \vdots & \vdots & \vdots & \vdots & \vdots & \ddots \\
\end{bmatrix} , 
\ee
\begin{figure}
  \includegraphics[scale=0.3,angle=0,width=0.97\linewidth]{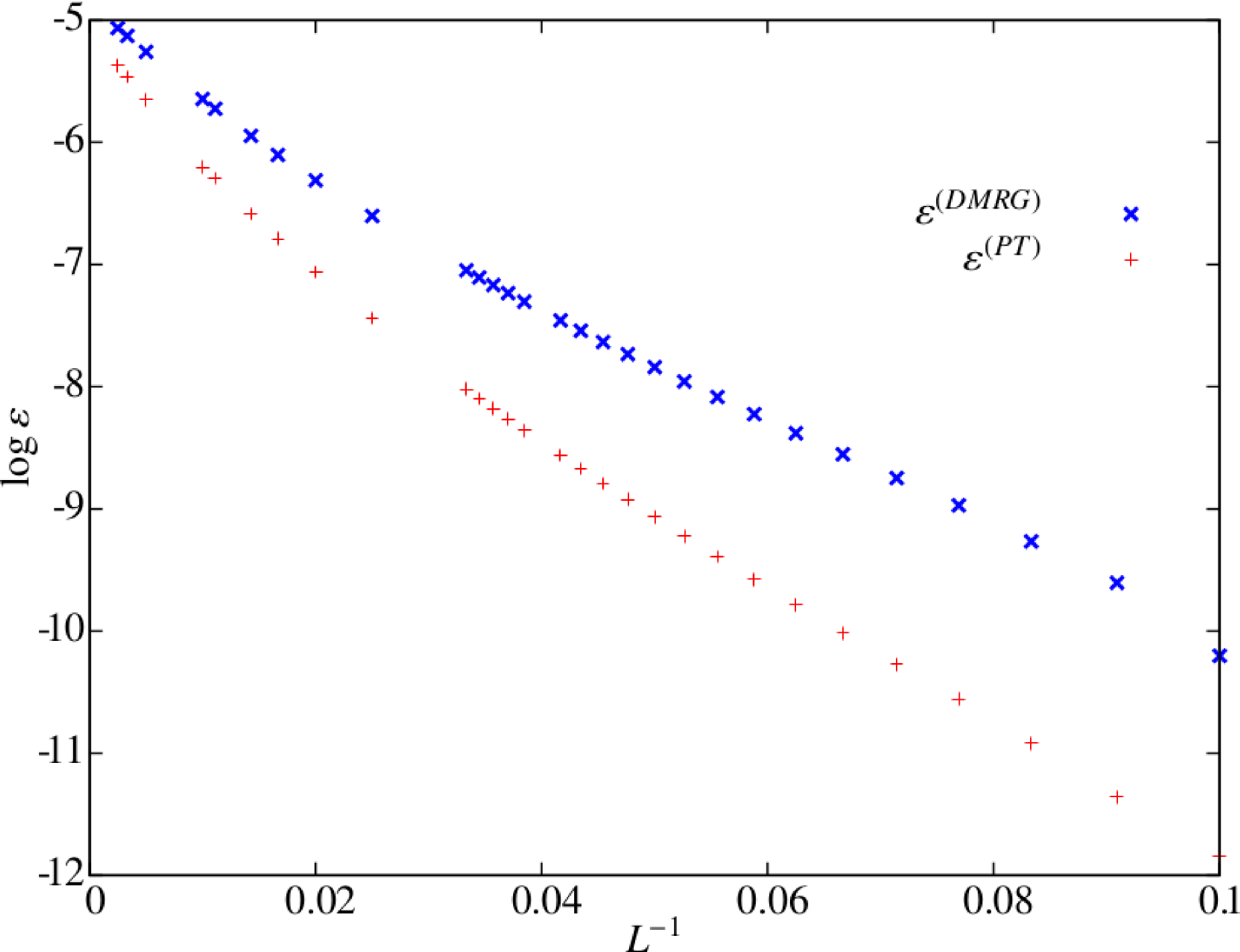}
  \caption{The error $\varepsilon$ of 1D Ising model as a function of length in the
quantum phase transition h = 1.0, for $\chi=8$.
We show that the error PT-DMRG In the large $L$ limit doesn't give a systematic improvement of accuracy.}
  \label{pt_vs_dmrg_strongL}
\end{figure}
where the non-zero elements are $a =
E_0$ at the top-left (the energy of the original ground state), a
series of $L$ entries along the top row and left column which is
$b=\langle \psi_i \vert H \vert \psi_0 \rangle$ (assumed independent
of i in the large $L$ limit), and the diagonal entries $c = \langle
\psi_i \vert H \vert \psi_i \rangle$ independent of $i$ in the large
$L$ limit. $a$ and $c$ are extensive in the system size, but $c$ has a
constant offset because of the local perturbation. So we can set:
\be
a = E_0 \times L \qquad c = E_0 \times L + q ,
\ee
where $q$ is the energy of the perturbation. It is
possible to determine the eigenvalues of this matrix as a function of
L, which is
\be
E = (E_0 \times L + \frac{q}{2}) - \Delta, 
\ee
where
\be
\Delta^2 = \frac{q^2}{4} + b^2 L .  
\ee
So we can see the origin now of the
$1/\sqrt(L)$ behaviour. For large $L$ the energy per site scales as
\be
\frac{E}{L} = E_0 - \frac{ \vert b \vert}{\sqrt{L}} + O(1/\sqrt{L}) .  
\ee
But in order to see the square root
behaviour $b^2 L >> q^2/4$, which for the ising model, requires L >
650 (see Fig. \ref{pt_vs_dmrg_strongL} ). The plot in Fig. 10 is basically linearizing a square root in a
region well away from the asymptotic large $L$ behaviour.
\section{Summary and outlook}

A simple and efficient numeric approach named PT-DMRG is proposed to
largely improve the accuracy of the conventional DMRG. It gives a
better approximation of ground state of strongly-correlated many-body
systems by recovering the leading term of entanglement that is
discarded in the truncations of DMRG. By using MPS representation, we
introduce a set of orthogonal basis to define the perturbed
Hamiltonian, whose ground state possesses a better precision than the
traditional DMRG. In other words, we use the Schmidt numbers that are
beyond the dimension cut-off to define the perturbation terms. By
using the second order PT-DMRG, our numerical results obtained for the
1D quantum Ising model and Heisenberg model show a better accuracy
reached by our PT-DMRG, where the precision of DMRG is shown to be
improved significantly (around $O(10)$ times).

Our PT-DMRG provides a fundamental scheme that can be directly used
for 2D DMRG algorithm. Such perturbation theory based on MPS can be
generalized to other MPS or even TN algorithms, such as time-evolved
block decimation. The generalization to higher-order perturbation
theories is be explored in the future.

Finally, the perturbation theories can provide a fundamental scheme to
study the power-low correlation. For example in the MERA the
isometries can be used to define perturbed terms. The kernel space of
each original isometry provides the tangent space in a natural way. So
the perturbation idea may be useful in any state ansatz that gives a
renormalization flow.
  
\section{Acknowledgements}

This work was supported by ERC ADG OSYRIS, EQuaM (FP7/2007-2013 Grant
No. 323714), Spanish MINECO (Severo Ochoa grant SEV-2015-0522), FOQUS
(FIS2013-46768), Catalan AGAUR SGR 874, Fundaci\'o Cellex and EU
FETPRO QUIOC, Marie Curie fellowship SQSNP 622939 FP7-MC-IIF.

\end{document}